\documentclass[epj]{svjour}
\usepackage[dvipsnames]{xcolor}
\usepackage{amsmath,amssymb}
\usepackage{bm}
\usepackage{amssymb, graphics, setspace}
\usepackage{pgfplots}
\usepackage{textcomp}
\usepackage[caption=false]{subfig}

\usepackage{amsmath}
\usepackage{commath}

\usepackage{graphicx,bm}
\usepackage{url}
\usepackage{float}
\usepackage{textcomp}
\usepackage{verbatim}
\begin{document}
\title{Entropy Production During Hadronization of a Quark-Gluon Plasma}
\titlerunning{QGP hadronization}
\author{ 
Tam\'as S. Bir\'o\inst{1} 
\and 
Zsolt Schram\inst{2} 
\and
L\'aszl\'o Jenkovszky\inst{3}
}
\authorrunning{T.S.~Bir\'o \and Z.~Schram \and L.~Jenkovszky}
\institute{%
H.A.S. Wigner Research Centre for Physics, Budapest, \email{Biro.Tamas@wigner.mta.hu}
\and %
Department of Theoretical Physics, University of Debrecen, Debrecen, \email{schram@phys.unideb.hu}
\and %
Bogoljubov Institute for Theoretical Physics, U.N.A.S. Kiev, \email{jenk@bitp.kiev.ua}
}
\date{Version: \today / Received: date / Revised: another date}
\abstract{%
We revisit the physical pictures for the hadronization of quark-gluon plasma,
concentrating on the problem of entropy production during processes where the
number of degrees of freedom is seemingly reduced  due to color confinement.
Based on observations on Regge trajectories
we propose not having an infinite tower of hadronic resonances.
We discuss possible entropy production mechanisms far from equilibrium
in terms of stochastic dynamics.
}

\maketitle

\newcommand{\be}{\begin{equation}}
\newcommand{\ee}[1]{ \label{#1} \end{equation}}

\newcommand{\ba}{\begin{eqnarray}}
\newcommand{\ea}[1]{ \label{#1} \end{eqnarray}}
\newcommand{\nl}{\nonumber \\}

\newcommand{\exv}[1]{ \left\langle {#1} \right\rangle }

\renewcommand{\Im}{\, \mathfrak{Im}\, }
\renewcommand{\Re}{\, \mathfrak{Re}\, }

\renewcommand{\pd}[2]{ \, \frac{\partial {#1}}{ \partial {#2}} \, }
\newcommand{\req}[1]{eq.(\ref{#1})}
\newcommand{\eon}[1]{ {\rm e}^{#1} }
\newcommand{\ave}[1]{ \left\langle {#1} \right\rangle }
\newcommand{\sumi}[1]{ \sum_{{#1}=0}^{\infty}\limits \,}


\section{Statistical Hadronization}

\textcolor{Red}{
}

Statistical models for describing the hadron and quark matter occurring in one or the other phase of a relativistic
heavy ion collision have several decades long history. Although all approaches agree with the experimental fact that
in the final state of such reactions only a plethora of different hadrons can be detected, they differ in the
physical picture put behind. In particular, since observable hadrons are not elementary particles,
any statistical theory has to deal with the reduction of the number of degrees of freedom due to
constructing complex bound states of quarks and gluons. However, in spontaneous processes in energetically
closed systems we do not expect a reduction in entropy, according to the second law of thermodynamics
\cite{Rafelski:Melting}.

Disregarding models which treat hadrons as fundamental constituents, the confinement of quarks and gluons into
hadrons poses the above outlined problem. There has been a nice variety of suggestions for circumventing this
trap. The early dynamical view considers tree graph level Feynman amplitudes producing practically enough
gluons, or hadrons, mainly pions, from predecessor quarks and antiquarks \cite{Klevansky,KlevanskyRehberg}. 
This helps the balance in
the number of degrees of freedom, but the color degeneracy has to be diminished at the end.
A simple fusion into a single, ground state hadron can produce a disaster for the entropy.

Combinatoric models, operating with constituent quark construction of final state hadrons,
also face this problem: although starting with massive, dressed quarks, these would have to be
in a color neutral state. Physical arguments to find a way out of this dilemma
include two main ingredients: i) a drastic volume expansion during the hadron formation,
ii) looking for a spectrum of ''excited'' constituent quark states, regarded as hadronic resonances.
Due to spin and isospin degeneracy, adding the first excited baryon octet and decuplet and the meson nonet
to the lowest quark bound states proved to be sufficient to accommodate experimental data without
an entropy reduction when comparing ideal gases on the quark-gluon plasma and hadron sides \cite{ALCOR,ALCORSPS}.

Beyond that it is a general problem that having an infinite number of possible hadronic resonance states
below a certain temperature, and only quark-gluon states above, the entropy and hence the pressure in the
hadronic phase is always larger \cite{Jakovac,BiroJakovac,JakoEPJA}. Therefore one cannot simply replace the infinite
number of hadron types by a restricted number of parton families; the melting of hadrons happens via the widening
and eventually merging of spectral densities to a continuum.

In this paper we investigate the possibility of having only a finite number of hadronic resonances,
derived from observations on Regge trajectories. Following this analysis we enlist statistical arguments
in favor of fluctuating numbers of hadronic degrees of freedom. It has observable consequences
for single hadron $p_T$ spectra obtained in high energy hadronic collisions.


\section{Hagedorn distribution and Regge Trajectories}

\textcolor{Red}{
}

Hadronic matter melts to quark matter rather like butter, not like ice \cite{AokiFodor}. At zero baryon density there is
no latent heat, proposing that the number of degrees of freedom change gradually and the temperature
is not kept constant during the transition process. Also at the reverse process, at the hadronization
of the quark gluon plasma into a hadron resonance gas the associated entropy production must measure
the change in the number of degrees of freedom. An assumed vigorous volume expansion  
is not fully compensated by cooling if dissipative effects are present.
It is therefore essential to understand how many hadron resonances are possible at all. Is the mass spectrum
of resonances really exponentially growing up to the infinite mass leading to the limiting Hagedorn temperature?
Or there is a saturation in this growth in number?

In order to seek answers to such questions we calculate the Hagedorn spectrum, Eqs.(\ref{Eq:H1},\ref{Eq:th}) based on the slope 
of the non-linear meson trajectories $\alpha^{\prime}(m)$,  via Eqs.(\ref{Eq:BK},\ref{Eq:A}), 
fitted to data on meson resonances. Our idea is to identify the prefactor $f(m)$ 
in Eq.(\ref{Eq:H}) with the derivative of the real part of non-linear Regge trajectories, 
$\Re\alpha^{\prime}(m)$, Eq.(\ref{Eq:H1}).

The main goal of this study is 
to understand the possible onset of saturation (termination of resonances).
Nevertheless, since at high masses the resonances tend to disappear gradually 
with decreasing height of the peaks and with increasing widths, 
it is not straightforward to conclude whether we see a smooth or a rapid transition
from the hadronic resonance gas to a quark-gluon plasma. 
Our combined analysis of both the Hagedorn distribution and the Regge trajectories 
provides a chance to discriminate between slow and sudden resonance melting dynamics.     

According to Hagedorn's conjecture \cite{H}, confirmed by subsequent studies \cite{H1,BF,BF1,BF1a,BF1b,BF1c,CK},
\begin{figure}[ht] 
\center{\includegraphics[width=.8\linewidth]{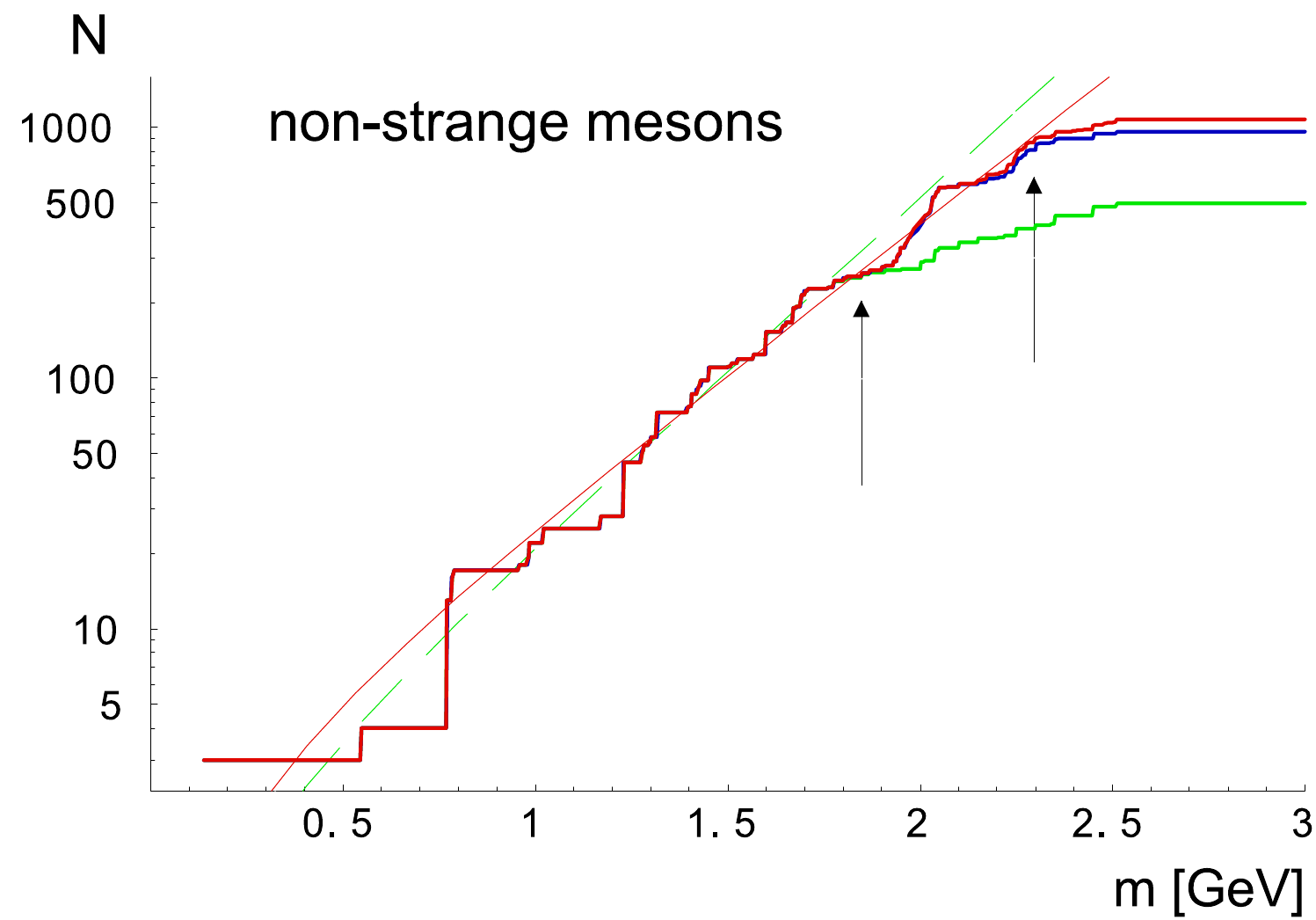} }
\caption{Accummulated spectrum of non-strange mesons plotted  as a function of mass. For more details see \cite{BF}}.  
\label{Fig:H}
\end{figure}
the density of hadronic resonances increases exponentially, supplemented by a slowly varying function of the mass, $f(m)$: 
\begin{equation}\label{Eq:H}
\rho(m)=f(m)\exp(m/T_H).
\end{equation}  
This is observed up to about $m=2\div 2.5$ GeV, whereupon the exponential rise slows down, see Fig.\ref{Fig:H} \cite{BF}. 
The temperature $T_H$, in the pre-QCD period called ´´critical´´ by Hagedorn, is nowadays identified with the color
deconfinement temperature. 
Alternatively, $T_H$ may not be connected with any temperature, being just a free parameter. 

Note that the above exponential form of mass distribution is not unique. As argued in \cite{Shuryak,Burak,BurakA,BurakB}, 
the density of states of a strongly interacting hadronic gas could be cubic in mass \cite{Landau}, 
$\rho(m)=4N(\alpha^{\prime})^2m^3$,
where $N$ is the finite number of parallel Regge trajectories and $\alpha^{\prime}$ is their universal slope.

Papers on resonance spectra (see e.g. \cite{BF1,BF1a,BF1b,BF1c,WBr} and references therein) may be divided into two groups:
one is devoted to the derivation and foundation of the mass spectrum from statistical physics, QCD or
Regge-pole and dual string models, the other is devoted to comparison with experimental data 
leading either to proofs of or observing a deviation from the Hagedorn distribution. 
The difference between meson and baryon spectra, flavour dependence, possible role of glueballs, 
departure from Hagedorn's exponential behaviour at large masses are particularly investigated topics.     
In the present paper we envisage the possible termination of resonances at some mass, $m\approx 2.5\div 3$ GeV. 


Apparently, a similar behaviour is typical for Regge trajectories in the resonance region.  
Two seemingly different phenomena, namely the well-known flattening of the Hagedern exponential distribution \cite{H}, 
cf. Fig.\ref{Fig:H}, around $m\approx 2\div 2.5$ GeV \cite{BF} and the less familiar turn-over of Regge trajectories 
at the same resonance mass range may both indicate, although in different ways, the onset of quark 
deconfinement by melting of hadrons. 
Imposing an asymptotic upper bound on the real part of any Regge trajectory,
\begin{equation}\label{Eq:Upper}
\Re\alpha(s)\leq const.
\end{equation} 
it follows - using analiticity and duality - that the Regge trajectories flatten
and the growth of the number of resonances terminates \cite{DAMA}.
This termination phenomenon was associated with an "ionization point" in various dual models \cite{Coon1,Coon2}.  

Possible links between Hagedorn behavior and Regge trajectories appear in the statistical bootstrap and 
dual models \cite{Statboot,StatbootA}. According to these the prefactor $f(m)$ in Eq.(\ref{Eq:H}) depends on the slope of 
the relevant Regge trajectory, $\alpha^{\prime}$. 
Very often a linearity of Regge trajectories is assumed.
A connection to the equation of state (EoS) was elucidated in Ref.\cite{Burak3}.

In the forthcoming we generalize the Hagedorn formula by utilizing the slope of relevant 
{\em non-linear} Regge trajectories into (\ref{Eq:H}). 
Anticipating a detailed quantitative analysis, one may observe immediately that 
the flattening\footnote{Due to crossing symmetry one can use the (positive) variables $s$ or $t$ interchangeably
implying also $\sqrt s=m$.}
of $\Re\alpha(s=m^2)$, indicated in Figs.\ref{Fig:R&I} and \ref{Fig:Rho},
results in a drastic decrease of the relevant slope $\alpha^{\prime}(m)$ and in a corresponding change 
of the Hagedorn spectrum. We modify Eq.(\ref{Eq:H}) as 
\begin{equation}\label{Eq:H1}
\rho(m)\sim(\Re\alpha^{\prime} (m))\exp(m/T).
\end{equation}

It is a standard procedure to regard the first cumulant of the resonance number spectrum, 
defined as the number of states with mass lower than $m_i$ \cite{BF,BF1,BF1a,BF1b,BF1c,WBr,1}.
The experimental values $g_i, m_i$ deliver the quantity
\begin{equation}\label{Eq:exp}
N_{exp}(m) \: = \: \sum_i g_i \, \Theta(m-m_i),
\end{equation}    
with $g_i=(2J_i+1)(2I_i+1)$ spin-isospin degeneracies of the $i$-th state and $m_i$  
respective masses. The theoretical prediction is obtained as
\begin{equation}\label{Eq:th}
N_{theor} \: = \: \int_0^m\limits\!\rho_{theor}(m') \, dm',
\end{equation} 
with
\begin{equation}
\rho_{theor}(m)=f(m) \, \exp(m/T)
\end{equation}
and  $f(m)\approx A(m^2+0.25 {\rm GeV}^2)^{-5/4}$ \cite{WBr}. 
Alternative choices for this slowly varying function are possible. 

Even without going into details, just by observing Fig.\ref{Fig:H} 
one inspects that the exponential increase in the density of states slows down 
around melting at $m\approx 2.5\div 3$ GeV 
{\em due to the decreasing factor $\alpha^{\prime}(m)$} in Eq.(\ref{Eq:H1}).
Note that the increasing number of experimentally found meson and baryon resonance states 
results basically from the degeneracy of a given multiplet. 
Details of counting the states can be found in \cite{Burak1,Burak,BurakA,BurakB}. 

\subsection{Nonlinear, complex Regge trajectories}  

\begin{figure}[ht] 
\center{\includegraphics[width=.6\linewidth]{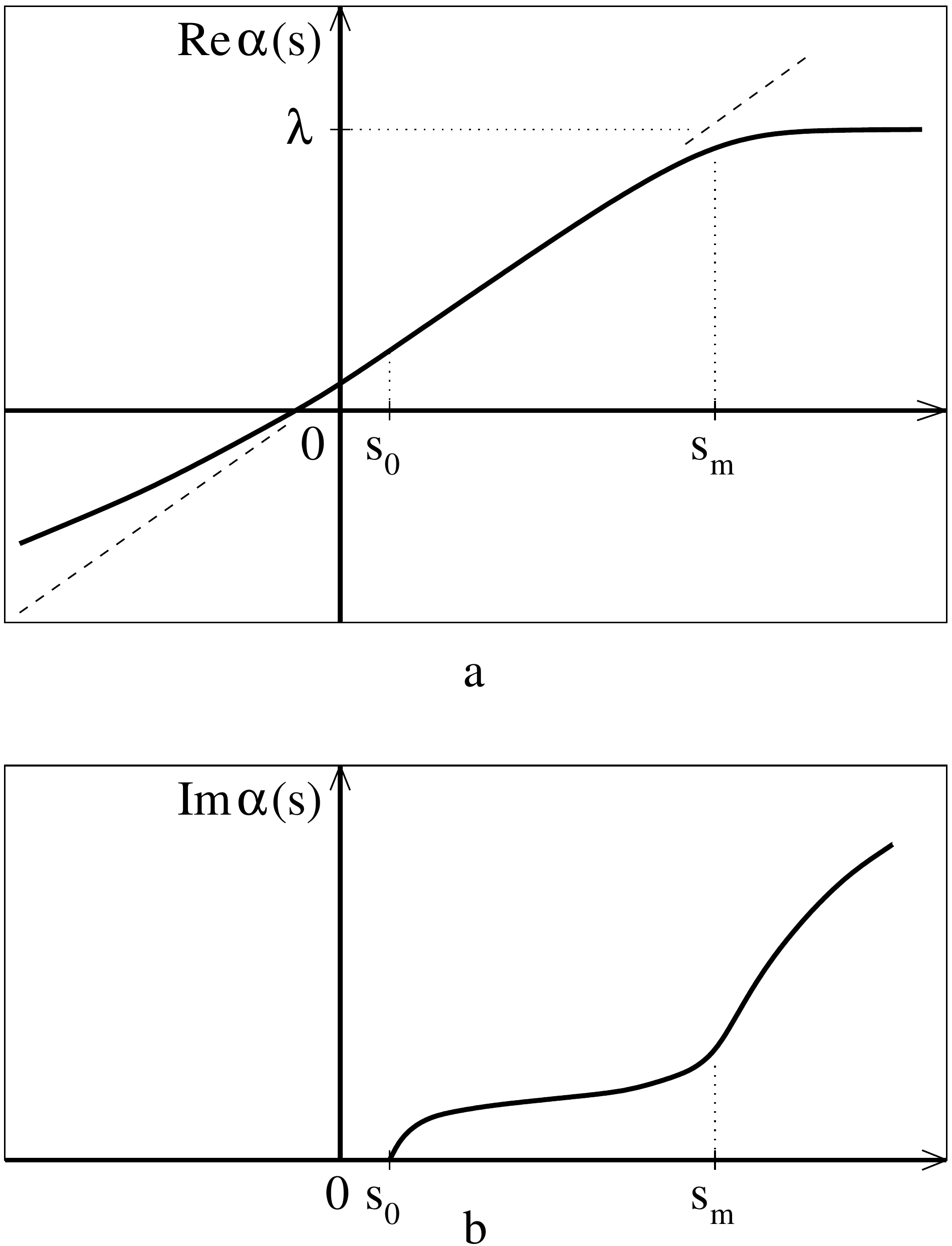} }
\caption{Typical behaviour of the real (a) and imaginary (b) parts of Regge trajectories in dual models with 
Mandelstam analyticity \cite{DAMA}}. 
\label{Fig:R&I}
\end{figure}

While the imaginary part of the Regge trajectories start from their first threshold, 
like $\sim{(s_0-s)^n}$, with $n\approx 1/2$ and $s_0=4m_{\pi}^2$,  
their real part is promoted by the highest threshold, $s_m$ (cf. Fig.\ref{Fig:R&I}). 
We do not know where exactly the resonance spectrum merges to a continuum, 
but have little doubt about the existence of 
an ''ionization point'', 
corresponding to a phase transition from melting hadrons to a boiling quark-gluon liquid, 
as predicted by lattice quantum chromodynamics (QCD) \cite{Rafelski:Melting}. 
The higher the threshold with the heaviest mass, $s_m=4m^2$, the further this citical point. 
In the limit of $m\to\infty$ one recovers a linear trajectory. 
In fact, it is more likely that the maximal $s_m$ is finite, belonging to the heaviest strongly 
interacting stable particles, e.g. bottomium, $s_m= 9.391$ GeV. 
The Barut-Zwanziger constraint on the threshold behavior, 
Eq.(\ref{Eq:BZ}), here does not apply.
We expect an asymptotic upper bound either following from DAMA/LIBRA experiment\cite{DAMA} 
or from more general considerations\cite{Tru}:
\ba
\mid\alpha(s)\mid_{\mid s\mid\rightarrow\infty}\leq O(\sqrt{\mid s\mid}), 
\nl \nl
\mid\alpha(s)/(\sqrt s\ln s)\mid_{\mid s\mid\rightarrow\infty}\equiv {\rm const} >0 
\nl \nl
{\rm or \quad} \Re\alpha(s)_{\mid s \mid\rightarrow\infty}\leq \rm{const}.
\ea{Eq:limit}
Wide-angle scaling behaviour of transition amplitudes and cross sections imposes an even more severe 
restriction, resulting in a logarithmic asymptotic behaviour of the trajectories,
\be
\alpha(s)\sim\ln(-s) \quad \mathrm{for} \quad |s| \rightarrow\infty.
\ee{Eq:ALFALOGES}
Using this asymptotics we smoothly interpolate between the low-$s$ linear behaviour 
and the logarithmic one. 
Moreover we also accomodate the square-root threshold in the form 
\be
\alpha(s)=\alpha_0-\sum_i\alpha_{1i}\ln(1+a_{2i}\sqrt{s_i-s}). 
\ee{SQUAREROTHTHR}
An even more stringent logarithmic bound may not affect the square-root trajectory
whose real part terminates at the highest-mass threshold anyway, but it will constrain its rise. 
By expanding the high-mass threshold in a power series, we get 
\begin{equation}\label{Eq:Tr1}
\alpha(s)=\alpha_0+\alpha^{\prime}s-\alpha_1\sqrt{s_0-s},  
\end{equation}
whose linear part should be slowed down asymptotically according to the asymptotic constraints.
It either happens by restoring the threshold or by imposing the logarithmic behaviour (\ref{SQUAREROTHTHR}). 
The existence of a maximally heavy threshold implies abrupt termination of resonances, corresponding 
to the deconfinement phase transition or termination of the Hagedorn distribution $\rho(m=\sqrt{s}).$ 
Termination of resonances, associated with an ''ionization point'' was studied also in a different class of dual models, 
based on logarithmic trajectories \cite{Coon1,Coon2}.  

Although everybody is aware of the non-linear nature of complex Regge trajectories, 
in most of the papers their linear approximation is used. The reason for this may be manifold:
i) The observed spectra of meson and baryon resonances (Chew-Frautchi plot) are very close to
linear functions; 
ii) the Veneziano-duality and hadronic string models imply linear Regge trajectories; 
iii) linearity simplifies calculations.
It is by far not a trivial problem to combine the nearly linear real part 
(spectrum of masses) with the highly non-linear imaginary part, connected to finite resonance widths. 
Attempts in this direction date back to the papers \cite{FP} and
\cite{P1,P2}, where the real and imaginary parts of the trajectories were connected by 
dispersion relations. In more pragmatic approaches \cite{BK1,Burak,BurakA,BurakB} 
explicit models compatible with theoretical constraints have been suggested and compared to the data. 

Let us recall here the main constraints on the trajectories imposed by the theory. 
First, the threshold behaviour of the Regge trajectories is constrained by unitarity. 
As shown by Barut and Zwanziger \cite{BZ}, unitarity constrains the Regge trajectories near their 
thresholds
$s\rightarrow s_0$ by
\begin{equation} \label{Eq:BZ}
\Im \alpha(s)\sim (s-s_0)^{\Re\alpha(s_0)+1/2},
\end{equation} 
where $s_0$ is the lowest threshold, say $4m_{\pi}^2$ in the case of the meson trajectories. 
Since $\Re\alpha(4m_{\pi}^2)$ is small, a square-root threshold is a reasonable approximation. 
High\-er thresholds, promoting the rise of the trajectories, may also be approximated by their power expansions. 
Second, as we mentioned earlier, in the resonance region $m=\sqrt{s}\lesssim 2.5$ GeV the meson and baryon 
trajectories are nearly linear (Chew-Frautchi plot). Asymptotically, the real part is constrained by
Eq.(\ref{Eq:BZ}). Fixed-angle scaling behaviour of the amplitude restricts the trajectories even more, 
down to a logarithm \cite{BCJ}. 

\begin{figure}[ht] 
\center{\includegraphics[angle=0, width=.7\linewidth]{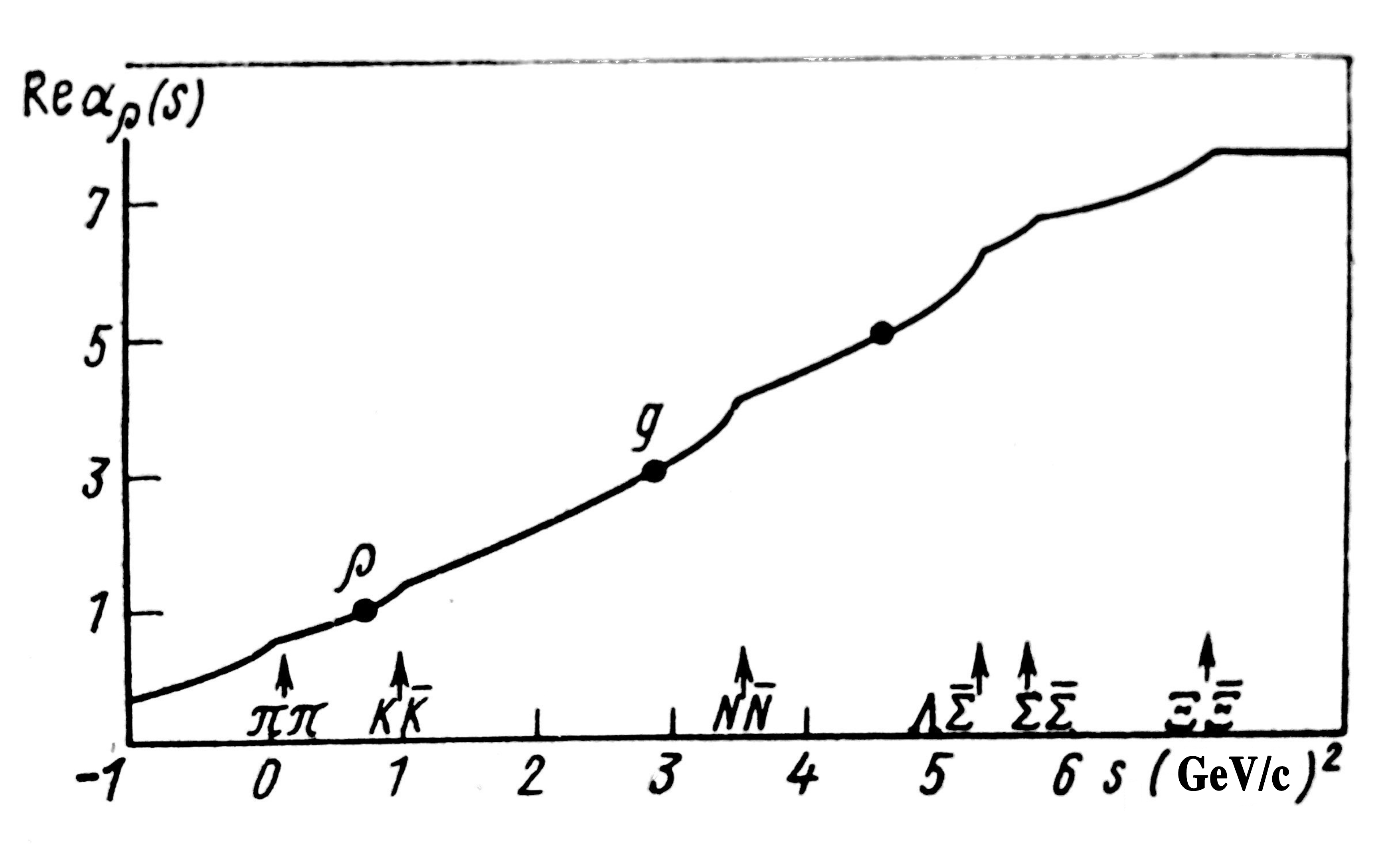} }
\caption{Model \cite{BK1} of the $\rho$ trajectory with $s_i=m_i^2$ thresholds indicated.} 
\label{Fig:Rho}
\end{figure}

Trajectories satisfying the above conditions and yet fitting the observed spectrum of resonances (cf. Fig.\ref{Fig:Rho})
can be found in \cite{BK1,Burak,BurakA,BurakB}. 
More ambitious models \cite{FP,P1,P2} use dispersion relations to relate the real and imaginary parts  of the trajectories. 
In Ref.\cite{Central} the $f$ trajectory was calculated according to \cite{P1}, 
while for the Pomeron it was approximated by a sum of square-root thresholds as advocated by \cite{BK1} 
in their study of meson and baryon resonances.  



Let us have a look at some examples. We start with a toy model of a
non-linear trajectory: following \cite{BK1}, we 
describe a simple trajectory in which the (additive) thresholds are those 
made of stable particles allowed by the quantum numbers. 
For the $\rho$ meson trajectory these are: 
$\pi\pi,\  K\bar K,\  N\bar N,\  \Lambda\bar \Sigma,\  \Sigma\bar \Sigma,\  \Xi\bar \Xi$. 
The relevant trajectory with parameters quoted in Ref.\cite{BK1}
is then given by
\ba
\alpha_{\rho}(m^2) & = &7.64-0.127\sqrt{m^2-0.28}-0.093\sqrt{m^2-0.988}
\nl
 & - & 0.761\sqrt{m^2-1.88} -  "\Lambda\bar \Sigma,\  \Sigma\bar \Sigma,\  \Xi\bar \Xi".
\ea{Eq:BK}

In Fig.\ref{Fig:Rho} the masses under the square roots are as listed above, and the coefficients in front 
of the square roots were adjusted like in Ref.\cite{BK1}, i.e. to fit the observed spectrum of resonances 
on the $\rho$ trajectory. The units are in GeV$^{-1}$. 
The resonances terminate at the highest threshold, here at $2.63$ GeV, whereupon the real part of the trajectory is flat. 
Similar models for other meson and baryon trajectories can be found in Refs.\cite{BK1}.

The trajectory (\ref{Eq:BK}) is by far not unique; moreover, it shows several deficiencies.
The square-root form of the threshold, reasonable for the lowest threshold, is only very approximative for the rest.
Furthermore, the choice of the thresholds made of stable particle pairs is an interesting one, but still not unique.
An obvious extension of the model can be the inclusion of thresholds made of particles with further flavours \cite{BK1}.
Ultimately, to avoid the proliferation of thresholds, one may replace them by effective quark masses. 
Important quantities are the first threshold, opening the imaginary part of the trajectory and the heaviest one 
promoting the real part until its termination. 

In the following we present several realisations of the above ideas with their practical applications, 
in particular having the Pomeron trajectory in mind. 
In order to handle the problem of decreasing widths of resonances, $\Gamma(m)$, 
mentioned at the beginning of this section, we apply the perscription 
\begin{equation}\label{Eq:Width}
\Gamma(m=\sqrt{s})=\frac{\Im\alpha(s)}{\Re\alpha^{\prime}(m^2)\sqrt{s}}>0.
\end{equation}


\subsection{Regge Models} 

Here we investigate a model for the trajectory similar to Eq.(\ref{Eq:BK}),
but with the square root dependence being replaced by novel terms ensuring positivity
of the real part of the coupling.  
The trajectory is now described as
\begin{equation}\label{Eq:A}
\alpha(m^2)=\frac{a+bm^2}{1+\sum_i\limits c_i\sqrt{m_i^2-m^2}}.
\end{equation}
This model was used recently in \cite{Central} for the Pomeron (glueball) trajectory.
To start with, we consider the $\rho$ trajectory including four thresholds: $\pi\pi,\ \ K\bar K,\ N\bar N$ and one more, 
with the highest threshold $s_4=s_H$ as a free parameter. We fit the parameters $a,\ b,\ c_i$ and $s_H$ 
to masses and decay widths of the resonances lying on the $\rho$ trajectory, and compare the resulting fit to the model 
described by Eq.(\ref{Eq:BK}). Note that in Ref.\cite{BK1} 
only the masses were involved in the fit. 
Here, apart from the masses, we include also the resonances' widths, calulated from Eq.(\ref{Eq:Width}). 
Data are taken from Ref.\cite{1}. 
Similar fits for other mesonic and baryonic trajectories were performed in Refs.\cite{P1,P2} and, 
more recently, in \cite{Diff} where a hypothetical glueball trajectory was also suggested.

We are particularly interested in the onset of saturation, where the rise of the real part of the trajectory terminates. 


To summarize, we have analysed three options: 

\begin{enumerate}
\item[I.]
Sum of square roots: this model, Eq.(\ref{Eq:BK}) with many thresholds of stable particle pairs,
was studied in detail for meson and baryon trajectories in Refs.\cite{BK1}. 
In Ref.\cite{Central} this model was applied with two thresholds, a light and a heavy one, 
and it was used for the Pomeron. As already said, it results in progressively decreasing widths of glueball resonances. 
This deficiency was cured in Ref.\cite{Diff} by means of a trajectory listed below (see III). 
In any case, within this option, adding at least one more threshold may be appropriate. 
 
\item[II.] 
The above model may be simplified by expanding only the highest threshold in a power series:
 \begin{equation}
 \alpha(s)=\alpha_0+\alpha' s-\alpha_1\sqrt{s_1-s}. 
  \end{equation}
Such a trajectory is particularly convenient in the small-$t$ scattering region, 
where the forward cone is known to be exponential with a ''fine structure'' due to the above-mentioned 
threshold singularity superimposed. This phenomenon is a hot topic by now, since it was confirmed by the 
TOTEM collaboration at the LHC (for details see \cite{Szanyi} and earlier references therein). 
In the particle region ($s>0)$ it is less attractive, for it predicts an infinite number of heavy resonances, 
lying on an almost linear trajectory. 

\item[III.] Eq.(\ref{Eq:A}) with the possibility to add more and more thresholds.
This trajectory is promising in resolving the problem of decreasing decay width of resonances \cite{Diff}. 
In fact, Eq.(\ref{Eq:Width}) was solved in Ref.\cite{Diff}, resulting in increasing widths of resonances, as required.

\end{enumerate}

To summarize this review, model I is attractive for its simplicity and flexibility in handling its free parameters. 
It results in a sharp cut (upper limit) of the resonance spectrum, compatible with the expectations from quantum chromodynamics. 
However, it also predicts a progressive narrowing of resonances. Option II is a simplified version of I, although 
with an infinitely rising spectrum of resonances. The model III may combine the required termination of resonances with 
a rising spectrum in a smooth way. The resonance widths in this case increase with the mass, a more natural 
scenario for hadron melting.


\section{Varying Phase Space Dimension}


In the usual treatment of canonical statistics it is assumed that the total energy of a microcanonical
distribution is fixed and a subsystem with small energy is investigated. In this case the subsystems'
energy fluctuates (for ideal gases it can even be analytically calculated to follow an Euler beta
distribution) and average properties define the temperature. In the limit of large total systems
a statistical weight, $\exp(-\epsilon/T)$ emerges as a factor in the probability of having
$\epsilon$ energy in the small subsystem.

Following this classical derivation of the thermodynamical temperature as $1/T=S^{\prime}(E)$, we
step back from this limit and consider cases when the total system is not large, but shows some
special features, possibly leading to another statistical weight than the exponential Boltzmann--Gibbs
factor. In this theoretical study we consider a simple phase space, where the volume
of all states with total energy $E$ and $n$ degrees of freedom is proportional to the $n$-dimensional
hypervolume, $\Omega(E) \sim E^n$. 

For different kinetic energy formulas for the individual particles the $n$-particle
phase space has different volumes and surfaces. We explore the consequences of a
phase space with fluctuating dimensions to the single particle energy spectra.
In \cite{Entropy16} and \cite{SQM16} we have considered the phase space volume of $n$ particles with a
total energy of $E$ as $\Omega(E)\propto E^n$. In a more realistic scenario we
explore now the $n$-dependence of possible coefficients.

Extreme relativistic particles moving in one dimension, as hadrons made
inside an energetic jet do, have kinetic energies of $\epsilon_i=|p_i|$. The phase space
volume of $n$ momentum components representing a total energy less than $E$, i.e. having momenta
between $-E$ and $E$ is given by the constraint
\be
 \sum_{i=1}^n\limits |p_i| \: \le \: E.
\ee{CONSTRAINT}
This is a $p=1$ case for a general $L_p$-norm:
\be
 \left( \sum_{i=1}^n\limits |p_i|^p \right)^{1/p} \: \le \: R(E),
\ee{LPCONSTRAINT}
with an energy dependent radius $R(E)$ in phase space.
The volume of the $L_p$-norm ball in $n$ dimension with radius $R$ is given by
\be
 \Omega_n^{(p)}(R) \: = \: \frac{\Gamma(\frac{1}{p}+1)^n}{\Gamma(\frac{n}{p}+1)} (2R)^n.
\ee{LPBALL}
So far we have considered the $p=\infty$ normed case,
\be
 \Omega_n^{(\infty)}(E) \: = \: (2E)^n,
\ee{LINFBALL}
being the volume of an $n$-dimensional hypercube.
For extreme relativistic (massless) particles the $p=1$ norm is relevant, having a hypervolume of
\be
 \Omega_n^{(1)}(E) \: = \: \frac{1}{n!} \, (2E)^n.
\ee{L1BALL}
This is the $n$-dimensional generalization of the diamond shape for $n=2$.
For massive, nonrelativistic particles in two dimensions, the $n$-ball is defined by a constraint
using the $p=2$ norm
\be
  \left( \sum_{i=1}^{2n}\limits p_i^2 \right)^{1/2} \: \le \: \sqrt{2mE}.
\ee{2DIMNBALL}
with $2n$ momentum component entries. The volume is given by
\be
 \Omega_{2n}^{(2)}(\sqrt{2mE}) \: = \: \frac{\Gamma(\frac{1}{2}+1)^{2n}}{\Gamma(\frac{2n}{2}+1)} \left(2\sqrt{2mE} \right)^{2n}
 \: = \: \frac{1}{n!} \, \left(2\pi m E \right)^n.
\ee{VOLTNDIM}
We repeat here the chain of thoughts already applied to the $p=\infty$ case earlier \cite{Entropy16,SQM16}.
Since we are interested in the statistical weight
of dividing the energy $E$ into a single particle energy $\epsilon$ and the rest $E-\epsilon$
we consider the ratio
\be
 r_n \: = \: \frac{\Omega_1(\epsilon) \, \Omega_{n-1}(E-\epsilon)}{\Omega_{n}(E)}.
\ee{RATION}
The phase space volume ratio in the $p=1$ norm is given as
\be
 r_n^{(1)} \: = \: \frac{\Omega_1^{(1)}(\epsilon) \, \Omega_{n-1}^{(1)}(E-\epsilon)}{\Omega_n^{(1)}(E)} \: = \:
 n \, \frac{\epsilon}{E} \, \left(1-\frac{\epsilon}{E} \right)^{n-1}.
\ee{RN1NORM}
The $p=2$ norm case based on eq.(\ref{VOLTNDIM}) 
results in the same expression as \req{RN1NORM} above:
\ba
 r_n^{(2)} \: &=& \: 
\frac{\Omega_2^{(2)}\left(\sqrt{2m\epsilon}\right) \, \, \Omega_{2(n-1)}^{(2)}\left(\sqrt{2m(E-\epsilon)}\right)  }{\Omega_{2n}^{(2)}\left(\sqrt{2mE}\right)} 
\nl
 \: &=& \: n \, \frac{\epsilon}{E} \, \left(1-\frac{\epsilon}{E} \right)^{n-1}.
\ea{RN2NORM}
For comparison the $p=\infty$ (hypercubic) result is
\be
 r_n^{(\infty)} \: = \: \frac{\epsilon}{E} \, \left(1 - \frac{\epsilon}{E} \right)^{n-1}
\ee{RINFTY}
It is easy to recognize that the result \req{RN1NORM} and \req{RN2NORM} can be written as a derivative
with respect to $\epsilon$:
\be
 r_n^{(1,2)} \: = \: - {\epsilon} \pd{}{\epsilon} \, \left(1-\frac{\epsilon}{E} \right)^n 
\ee{R1DERIV}
Two important examples are here to be discussed. The first is a Poisson-distributed
number of particles,
\be
 P_n \: = \: \frac{\exv{n}^n}{n!} \eon{-\exv{n}},
\ee{POIS}
characteristic to Glauber coherent states of bosons.
This delivers
\be
 \exv{r_n^{(1,2)}}_{{\rm POI}} \: = \: \frac{\epsilon}{T} \, \eon{-\epsilon/T}
\ee{POIR1}
with $T=E/\exv{n}$.
In the second example the number of hadrons is distributed according to a negative binomial
distribution,
\be
 P_n \: = \: \binom{n+k-1}{n} \, f^n \, (1+f)^{-n-k}.
\ee{NBD}
This follows from a $k$-fold Bose statistics and can also be viewed as a non-Glauber (''nonlinear'')
optical coherent state. The expectation value of the phase space ratio becomes
\be
 \exv{r_n^{(1,2)}}_{{\rm NBD}} \: = \: \frac{\epsilon}{T} \, \left(1+\frac{\epsilon}{kT} \right)^{-k-1}.
\ee{R1NBD}
With $q=1+1/k$ this resembles a Tsallis-Pareto distribution as the factor besides the one-particle phase space
volume for energy $\epsilon$:
\be
 \exv{r_n^{(1,2)}}_{{\rm NBD}} \: = \: \frac{\epsilon}{T} \, \left(1+(q-1)\frac{\epsilon}{T} \right)^{-\frac{q}{q-1}}.
\ee{R1NBDTSALLIS}
In the case of a general $n$-distribution, $P_n$, the consideration is based on the average statistical factor
\be
 \overline{\rho} \: := \: \sumi{n} \left(1- \frac{\epsilon}{E} \right)^n \, P_n,
\ee{AVSTATFAC}
used in \req{R1DERIV}.
While in the following section we shall concentrate on hadronization ideas dealing with the production
and properties of the NBD, here we just give a brief review of the physical interpretation
of the spectral parameters $T$ and $q$ for general distributions and its connections to 
thermodynamics.  They cannot represent a Tsallis--Pareto result exactly, but they very often
represent it approximately. Expanding both the definition eq.(\ref{AVSTATFAC}) and the Tsallis--Pareto
result in $\epsilon$ up to second order, one concludes that
\be
 \frac{1}{T} \: = \: \frac{\ave{n}}{E} \qquad {\rm and} \qquad \frac{q}{T^2} \: = \: \frac{\ave{n(n-1)}}{E^2}.
\ee{PARAMS}
This agrees with the kinetic interpretation of temperature in general, $T=E/\ave{n}$, and also
gives an interesting interpretation of $q$ as being the second scaled factorial moment:
\be
 q \: = \: \frac{\ave{n(n-1)}}{\ave{n}^2}.
\ee{QASMOMENT}
In other words $q \ne 1$ reflects the non-Poissonity of the $P_n$ distribution.

A connection to thermodynamical formulas can be established, if we abandon the picture based  solely
on the fluctuation of the number of degrees of freedom, $n$, and extend our investigations
towards a system described by the entropy--energy function, called an equation of state.
In this case the canonical statistical factor averaged over repeated independent experiments
becomes
\be
 \overline{\rho}(\epsilon) \: = \: \ave{\eon{S(E-\epsilon)-S(E)}},
\ee{RHOEINSTEIN}
following Einstein's assumption about the fundamental relation between the occupied phase phase volume
and the entropy. Again, expansion up to ${\cal O}(\epsilon^2)$ terms delivers
\be
 \frac{1}{T} \: = \: \ave{S^{\prime}(E)} \qquad {\rm and} \qquad
 q \: = \: \frac{\ave{S^{\prime}(E)^2 \, + \, S^{\prime\prime}(E)}}{\ave{S^{\prime}(E)}^2}.
\ee{THERMOPARAMS}
Here the first result is the generalization of the temperature concept known from classical
thermodynamics as the average of the event-by-event fluctuating value, $1/T=\ave{\beta}$,
with $\beta=S^{\prime}(E)$ being the familiar definition. Following this correspondence
the long sought interpretation of the $q$ parameter connects it with the variance of
this quantity, and due to $\ave{S^{\prime\prime}(E)}=d(1/T)/dE=-1/CT^2$ with the total
heat capacity, $C$, of the system:
\be
 q \: = \: 1 + \frac{\Delta \beta^2}{\ave{\beta}^2} - \frac{1}{C}.
\ee{qTOTAL}
One realizes that a general $q$, approximately describing canonical distributions to
finite systems with fluctuating temperature, can both be smaller or larger than one.
In fact, the traditional picture {\em assumes} $q=1$ and concludes that the
temperature fluctations would satisfy $\Delta\beta/\ave{\beta}=1/\sqrt{C}$.
Since $C$ in extensive systems is proportional to the total number of degrees of freedom,
this is the origin of the ''one over square root'' suppression of finite size effects
commonplace statement.

This does not imply that the distribution of possible $\beta$ values were Gaussian,
any distribution with the same variance leads to the same $q$ parameter while approximating
the single particle energy spectra with the Tsallis--Pareto form. In fact our picture based
on the fluctuation of the number of degrees of freedom in the finally observed phase space volume
can easily be related to the idea of distributed $\beta$ values, also called {\em superstatistics}.
From the equality
\be
 \overline{\rho}(\epsilon) \: = \: \sumi{n} P_n \, (1-\epsilon/E)^n \: = \: \int\!w(\beta)\eon{-\beta E} \, d\beta,
\ee{SUPERSTAT}
it is sraightforward to derive that $P_n$ is the Poisson transform of the superstatistical
distribution:
\be
 P_n \: = \: \int \frac{(\beta E)^n}{n!} \eon{-\beta E} \, w(\beta) \, d\beta.
\ee{POITRF}
The underlying trick is to use the identity 
\be
\eon{-\beta\epsilon}=\eon{\beta E(1-\epsilon/E)}\eon{-\beta E}
\ee{}
and to identify the coefficient of $(1-\epsilon/E)^n$ after expanding the first exponential factor.
In particular the $w(\beta)=\delta(\beta-1/T)$ sharp superstatistics leads to the Poisson distribution
in $n$, while the NBD distribution is gained starting with a gamma distribution for the $\beta$
values. This superstatistical view, however, knows nothing about the interplay between finite heat
capacity and degrees of freedom fluctuations inherent in eq.(\ref{qTOTAL}).

We note 
that the Tsallis--Pareto distribution in $\epsilon$ may also be viewed as optimizing
an entropic variational principle using the Tsallis or Renyi entropy formula, different from the
classical Boltzmannian one. This approach also lacks a physical interpretation for the parameter $q$,
which has to be added. Our approach, outlined above, offers a natural and general interpretation
for this quantity.

\section{Unidirectional and Resetting Processes}

\textcolor{Red}{
}

We are now left with the problem to understand the emergence of negative binomial distributions
in the number of final state hadrons dynamically, with the hope that the observed (fitted)
values for $q$ parameters can then be connected with characteristics of physical mechanisms
present in (soft) hadronization. It would also be nice to understand in what extent a 
non-Boltzmannian entropy formula can be useful for describing such processes.
Following this aim we discuss a more general class of possible schemes of stohastic
dynamics \cite{BiroNeda}.


Let us consider the following family of equations:
\be
 \dot{P}_n \: = \: \sum_m\limits \left[w_{nm} a(P_m) - w_{mn} a(P_n) \right].
\ee{STOCDYNA}
Due to its antisymmetric construction in the indices $n$ and $m$ the conservation of probability
normalization is guaranteed, $\sum_n \dot{P}_n = 0$. Here $a(P)>0$ is a generic function, allowing for
nonlinear dependences on the occupation probability of the initial state during a transition from
one state to another. The stationary distribution, $Q_n$, satisfies:
\be
 0 \: = \: \sum_m\limits \left[w_{nm} a(Q_m) - w_{mn} a(Q_n) \right].
\ee{STACA}
We seek for an entropic distance definition in the trace form
\be
 \rho(P,Q) \: := \: \sum_n\limits  \sigma_n(P_n,Q_n) \: \ge \: 0,
\ee{ENTDISTA}
with the triviality property $\rho(Q,Q)=0$ and otherwise positive.
The evolution of this entropic distance is given by 
\be
\dot{\rho}=\sum_n \pd{\sigma_n}{P_n} \cdot \dot{P}_n,
\ee{RHODOT}
applying eq.(\ref{STOCDYNA}) it appears as
\be
 \dot{\rho} \: = \: \sum_{n,m}\limits \pd{\sigma_n}{P_n} \, \left[w_{nm}a(Q_m)\xi_m - w_{mn} a(Q_n) \xi_n \right],
\ee{RHODOTA}
using the notation $\xi_n = a(P_n) / a(Q_n)$. In the second term of the above double sum we can perform the summation
over $m$ using the stationarity condition (\ref{STACA}) and with arbitrary $\lambda_n$, $\lambda_m$ arrive at
\be
 \dot{\rho} \: = \: \sum_{n,m}\limits
 \left[\pd{\sigma_n}{P_n}(\xi_m-\xi_n) + (\lambda_n-\lambda_m) \right] w_{nm} a(Q_m).
\ee{RHODOT2A}
Here we utilize the stationarity eq.(\ref{STACA}) in order to show that the terms containing the arbitrary
factor $\lambda_n-\lambda_m$ sum up to zero. It is based on exchanging the summation indices $n$ and $m$
in the subtracted second term:
\ba
 \sum_{n,m}\limits (\lambda_n-\lambda_m) w_{nm} a(Q_m) \: &=& \:
\nl 
  \sum_n\limits \lambda_n \sum_m\limits \left[ w_{nm}a(Q_m) - w_{mn} a(Q_n) \right]  \: &= \: 0.
\ea{ITISZERO}
For satisfying the second law of thermodynamics one aims to attend $\dot{\rho}$ with a definite sign.
It is possible if and only if one constructs
\be
 \pd{\sigma_n}{P_n}  \: = \: \mathfrak{s}^{\prime} (\xi_n),
\ee{IMPORTANT}
and chooses $\lambda_n = \mathfrak{s}(\xi_n)$ in eq.(\ref{RHODOT2A}). Then the factor in the square brackets
becomes
\be
 \mathfrak{s}^{\prime}(\xi_n) \, (\xi_m-\xi_n) + \mathfrak{s}(\xi_n) - \mathfrak{s}(\xi_m) \: = \:
 - \frac{1}{2} \mathfrak{s}^{\prime\prime}(c_{mn}) \, \left( \xi_m - \xi_n \right)^2,
\ee{SQBFAC}
as an application of the remainder theorem for Taylor series in the Lagrange form. Here $c_{mn}$
is a value between $\xi_m$ and $\xi_n$, endpoints included. In this way our final result for the change
of the entropic distance reads as
\be
 \dot{\rho} \: = \: - \frac{1}{2} \sum_{n,m}\limits \mathfrak{s}^{\prime\prime}(c_{mn}) \,
 \left( \xi_m - \xi_n \right)^2 \, w_{mn} \, a(Q_m).
\ee{RHODOTFIN}
We conclude that the only requirement for $\dot{\rho} \le 0$ is the concavity of the function
$\mathfrak{s}(\xi)$, i.e. we require $\mathfrak{s}^{\prime\prime} > 0$ for all its possible arguments.
Having once such a function, $\mathfrak{s}(\xi)$, one reconstructs the entropic distance based on
$\sigma_n(P_n,Q_n)$ by solving the partial differential equation
\be
 \pd{\sigma_n}{P_n} \: = \: \mathfrak{s}^{\prime} \left(\frac{a(P_n)}{a(Q_n)} \right).
\ee{SOLVETHIS}
Integration constants in the solution are to be set ensuring $\rho(Q,Q)=0$.

To give a classical example, for linear dynamical models, $a(P)=P$, one uses $\mathfrak{s}(\xi)=-\ln \xi$
and arrives at the Kullback-Leibler divergence formula for the entropic distance.
Here we generalized the classical construction. Another example is given by the R\'enyi-divergence.
One uses $\mathfrak{s}(\xi)=(\xi^{-\nu}-1)/\nu$, having in this way $\mathfrak{s}^{\prime}(\xi)=-\xi^{-\nu-1}$
and $\mathfrak{s}^{\prime\prime}(\xi)=(\nu+1)\xi^{-\nu-2} > 0$, and considers nonlinear dynamics
with $a(P)=P^{\lambda}$. Using the notation $q=\lambda(\nu+1)$ one obtains
$\pd{\sigma_n}{P_n}=-(Q_n/P_n)^q$, and in this case the solution of eq.(\ref{SOLVETHIS}) 
with the proper integration constant leads to
\be
 \rho(P,Q) \: = \: \frac{1}{1-q} \left(1-\sum_n\limits Q_n^q P_n^{1-q} \right).
\ee{RENYIDIV}
Note that the uniform distribution, $U_n=1/N$ for $n=1,2,\ldots N$ has the entropic distance
\be
 \rho(U,Q) \: = \: N^{q-1} \left\{S_T(U) - S_T(Q)  \right\}
\ee{URENYIDIV}
with 
\be
 S_T(Q) \: = \: \frac{1}{1-q} \sum_n\limits \left( Q_n - Q_n^q \right)
\ee{TSALENTROP}
being the Tsallis entropy. Both its non-extensive property for $q \ne 1$ and the possible interpretation
of the divergence as relative information can be well received from this result.

Since the number of newly made hadrons fluctuates event by event and it is intimately connected
to non-perturbative physics effects, the distribution $P_n$ can be measured, but cannot be calculated
from first principles to date. To a very good approximation $P_n$ is a negative binomial distribution
(NBD) in $pp$, $pA$ and $AA$ collisions. The question for model makers arises how to obtain an NBD
possibly stable. Both type of models described in the following, namely a small jump model and also a big jump,
asymmetrically directed model is able to result in such a distribution as a stationary solution
of the underlying simplified stochastic dynamics.

We consider here first the simple combinatoric model with repetitive placement of hadrons (bosons)
in phase space cells. Then the number of equivalent arrangements of $N$ particles in $K$ cells
counts as
\be
 \Omega(N,K) \: = \: \binom{N+K}{N}.
\ee{COMBI}
For the observation of fluctuations one picks up $n$ hadrons in $k$ cells, while the remaining
$N-n$ hadrons are distributed evenly among $K-k$ cells. The resulting Boltzmannian probability
is the fraction of such special arrangements,
\be
 Q_n \: = \: \frac{\binom{n+k}{n} \, \binom{N-n+K-k}{N-n} }{\binom{N+K+1}{N}}.
\ee{POLYAQN}
This so called P\'olya distribution, originally gained in urn draw games with repositioned 
stones, in the large system limit, $N\to\infty$, $K\to\infty$ with fixed $f=N/K$
approaches the NBD distribution
\be
 Q_n \: = \: \binom{n+k}{n} \, f^{n} \, (1+f)^{-n-k-1}.
\ee{NBDQN}
Let us compare here two dynamical models with different philosophy, but both leading to the same
NBD as stationary distribution. In a picture when hadrons can be made and eliminated one by one,
only the near-diagonal transition rates differ from zero, $w_{nm}=\mu_m\delta_{m,n-1}+\lambda_m\delta_{m,n+1}$.
In this case the dynamical model reads as\footnote{For  $\lambda_n=\mu_n=\sigma$ this is a discrete model
of the Fokker--Planck equation, describing diffusion in the state space.}
\be
 \dot{P}_n \: = \: \lambda_{n+1}P_{n+1}-\mu_nP_n + \mu_{n-1}P_{n-1}-\lambda_nP_n.
\ee{ONEBYONE}
The stationary distribution satisfies $\lambda_nQ_n=\mu_{n-1}Q_{n-1}$, which for an NBD
as in eq.(\ref{NBDQN}) requires
\be
 \frac{f}{1+f} Q_{n-1} \: = \: \frac{n}{n+k} Q_n.
\ee{NNBDRECUR}
This in turn assigns the following state-dependent rates to the model:
\be
 \lambda_n = \sigma (1+f)n, \qquad {\mathrm{and}} \qquad \mu_n = \sigma f (n+k+1).
\ee{NBDRATES}
On the other hand a process which makes only new hadrons one by one, and diminishes
only the whole, has upward transition rates $w_{n+1,n}=\mu_n$ and a special rate
for annulation,  $w_{n,m}=\mu_m\delta_{m,n-1}+\gamma_m \delta_{n,0} $. 
The dynamical model which describes such processes is given by
\be
 \dot{P}_n \: = \: \mu_{n-1}P_{n-1} - (\mu_n+\gamma_n) P_n.
\ee{UNIDIR}
for $n \ge 1$ and by $\dot{P}_0=\exv{\gamma_n}-(\mu_0+\gamma_0)P_0$.
It is unidirectional and resetting, while it also has a stationary distribution resembling
the one in the first model: $\mu_{n-1}Q_{n-1}=(\mu_n+\gamma_n)Q_n$.
The necessary resetting rate can be obtained from the above result eq.(\ref{NBDRATES})
as
\be
 \gamma_n = \lambda_n - \mu_n \: = \: \sigma (n - f(k+1)).
\ee{ALTERNATIVRATES}
We note that in this case $\exv{\gamma_n}=0$ for the stationary NBD, and worse than that, for
low values of $n$ the resetting rate $\gamma_n=w_{0,n}$ would be negative. This invalidates
all of our earlier proofs for the positive entropy production (following from the reduction
of entropic divergence from the stationary distribution).

We conclude that a viable unidirectional scenario is imaginable only for a modified
NBD,
\be
 Q_n \: \propto \: \frac{\binom{n+\alpha+k}{n+\alpha}}{\binom{\alpha+k}{\alpha}} \, x^n
\ee{MODIFIED}
with suitable normalization. One obtains $\gamma_n > 0$, if $\alpha \ge f(k+1)$.
For small $f$ and moderate $k$ this can be a minor modification to the original NBD
and might reflect a viable dynamical model with positive rates and entropy production.

\section{Conclusion}

Flattening of the exponential density of states and that of the linear rise of Regge trajectories point to the same phenomenon, 
namely to quark deconfinement and the special way how hadrons melt. These two phenomena are correlated but they are not 
necessarily identical. Their combined study and subsequent fits to data may reduce the presently available freedom in the relevant 
parametrizations and shall tell us more about the unset of deconfinement. 


This paper is devoted to the memory of Walter Greiner. He was a man of strong character and led the Frankfurt
School of Research with eminence and keen instinct. His scope on nuclear, heavy ion and strong field physics
embraced international teams, supporting the collaboration also with ''the other Europe'' in times of
artificially created animosity between East and West. Among his ramified activities he has always supported
Hungarian physicists, in particular the ''Nuclear Troika'' (and their students), as he named them in a limerick
verse written by himself, in Acta Physica Hungarica.


With this dedication to our late Teacher, Colleague and Friend, we think we follow
the spirit of Walter Greiner's scientific heritage in two important
aspects: one is his universal search for harmony
in Nature - from atoms, through atomic nuclei to the smallest pieces of 
hadronic matter - quarks and gluons; the other one is his admiration for new, unorthodox
methods and phenomena as opposed to the comfortable and "safe" follow-up of the main stream.

In this paper we presented various approaches to some burning
problems in strong interaction dynamics, from low-energy resonances,
to high-energy multiple production of hadrons. In doing so, we proposed
among others an unorthodox approach to hadronic resonances, lying on
non-linear Regge trajectories (Chew-Frautchi plot), predicting a limited
number of resonances in Nature, which is contrary to the traditional view
based {\it e.g.} on string models and linear Regge trajectories, and
predicting an infinite number of resonances. Yet, we relate this model to
Hagedorn's statistical approach, recalling basic thermodynamical principles, 
unifying thus two seemingly distinct perspectives. We mean Walter would have liked it!


\begin{acknowledgement}
This work has been supported by the bilateral academic cooperation between the Hungarian Academy of Science
and the Ukrainian National Academy for Sicence under the mutual visitor exchange project 
{\em Supercooled metastable states in heavy ion collisions and compact stars}, and by
the Helmholtz International Center for FAIR within the LOEWE program launched by the State of Hesse.
L.~J. was also supported by the Ukranian Academy of Sciences' project {\em Matter under extreme conditions}.
Support, iniated from September 1, 2017 by the Hungarian National Bureau for Research, Development and Innovation,
NKFIH via the project  K 123815 is acknowledged.
\end{acknowledgement}

%


\begin{thebibliography}{}
%
%

\bibitem{Rafelski:Melting}
J.~Rafelski,
EPJ A {\bf 51} (2015) 114

\bibitem{Klevansky}
D.~S.~Isert, S.~P.~Klevansky,
EPJ A {\bf 12} (2001) 453

\bibitem{KlevanskyRehberg}
D.~S.~Isert, S.~P.~Klevansky, P.~Rehberg,
Nucl.~Phys. A {\bf 643} (1998) 275

\bibitem{ALCOR}
T.~S.~Bir\'o, P.~L\'evai, J.~Zim\'anyi,
Phys.~Lett. B {\bf 347} (1995) 6


\bibitem{ALCORSPS}
J.~Zim\'anyi, T.~S.~Bir\'o, T.~Cs\"org\H{o}, P.~L\'evai,
Phys.~Lett. B {\bf 472} (2000) 243


\bibitem{Jakovac}
A.~Jakov\'ac,
Phys.~Rev. D {\bf 88} (2013) 065012

\bibitem{BiroJakovac}
T.~S.~Bir\'o, A.~Jakov\'ac,
Phys.~Rev. D {\bf 90} (2014) 045038

\bibitem{JakoEPJA}
T.~S.~Bir\'o, A.~Jakov\'ac,
EPJ A {\bf 53} (2017) 52

\bibitem{AokiFodor}
Y.~Aoki, G.~Endr\H{o}di, Z.~Fodor, S.~D.~Katz, K.~K.~Szab\'o,
Nature {\bf 443.7112} (2006) 675


\bibitem{H} R.~Hagedorn, Nuovo Cim. Suppl. {\bf 3} (1965) 147

\bibitem{H1} 
J.~Cleymans, D.~Worku,
Mod.~Phys.~Lett. A {\bf 26} (2011) 1197
 
\bibitem{BF} 
W.~Broniowski, W.~Florkowski and L.~Glozman, 
Phys.~Rev. D {\bf 70} (2004) 117503


\bibitem{BF1} 
W.~Broniowski, W.~Florkowski, 
Phys.~Lett. B {\bf 490} (2000) 223

\bibitem{BF1a}
W.~Broniowski, E.~R.~Arriola, 
POS LC (2010) 2010:062


\bibitem{BF1b}              
E.~R.~Arriola, W.~Broniowski, P.~Masjuan,
ERA at Light Cone, Cracow, JUly 8-13, 2012. 
arXiv: hep-ph/1210.7153; 

\bibitem{BF1c}
W.~Broniowski, 
{\em Limits on hadron spectrum from bulk medium properties,}
arXiv: nucl-th/1610.0967


\bibitem{WBr} 
W.~Broniowski, 
{\em  Distinct Hagedorn temperatures from particle spectra: a higher one for mesons, a lower one for baryons,}
Few-quark problems, Bled, 8-15 July 2000,
arXiv: hep-ph/0008112

\bibitem{1} 
K.~A.~Olive et al. (Particle Data Group), Chinese Physics C{\bf 38} (2014) 090001 
 

\bibitem{CK} 
T.~D.~Cohen, V.~Krejcirik, 
J.~Phys. G (Nucl.Part.Phys.) {\bf 39} (2012) 055001

\bibitem{Landau} 
S.~Z.~Belenky, L.~D.~Landau, Sov. Phys. Uspekhi {\bf 56} (1955) 309

\bibitem{Shuryak} 
E.~V.~Shuryak, Sov.~J.~Nucl.~Phys. {\bf 16} (1973) 220


\bibitem{Burak1} 
L.~Burakovsky, 
PRQT ' 98 Conf. Houston, Texas, Feb 9-11, 1998,
arXiv: hep-ph/9805286


\bibitem{Burak2} 
M.~M.~Brisudova, L.~Burakovsky, T.~Goldman, A.~Szczepaniak, 
Phys.~Rev. D {\bf 67} (2003) 094016


\bibitem{Burak3} 
L.~Burakowsky, L.~P.~Horowitz, 
Nucl.~Phys. A {\bf 614} (1997) 373


\bibitem{Central} 
R.~Fiore, L.~Jenkovszky, R.~Schicker, 
EPJ C {\bf 76} (2016) 1-10 

\bibitem{FP} 
A.~Degasperis, E.~Predazzi, 
Nuovo Cim. {\bf A 65} (1970) 764 

\bibitem{P1} 
R.~Fiore {\it et al.},
EPJ. A {\bf 10} 217-221 

\bibitem{P2} 
R.~Fiore {\it et. al.},
Phys. Rev. D {\bf 70} (2004) 054003 

\bibitem{BZ} 
A.~O.~Barut, D.~E.~Zwanziger, 
Phys. Rev. {\bf 127} (1962) 974

\bibitem{BK1} 
A.~I.~Bugrij, N.~A.~Kobylinskij, 
Annalen d. Physik {\bf 32} (1975) 297



\bibitem{Burak} 
M.~M.~Brisudova, L.~Burakovsky, T.~Goldman, 
Phys.~Rev. D {\bf 61} (2000) 054013

\bibitem{BurakA}
M.~M.~Brisudova, L.~Burakovsky, T.~Goldman, 
Phys.~Lett. B {\bf 460} (1999) 1

\bibitem{BurakB}
L.~Burakovsky, T.~Goldman, 
Phys.~Rev.~Lett. {\bf 82} (1999) 457

 
\bibitem{DAMA} 
G.~Cohen-Tannoudji {\it et al.}, 
Fortschritte d. Physik, {\bf 21} (1973) 427

\bibitem{Tru} 
A.~A.~Trushevsky, 
Ukr.~Fiz.~Zh. {\bf 22} (1977) 353

\bibitem{BCJ} 
A.~I.~Bugrij, Z.~E.~Chikovani, L.~L.~Jenkovszky, 
Z.~Phys. C {\bf 4} (1980) 45

\bibitem{Coon1} 
D.~D.~Coon, 
Phys.~Lett. B {\bf 22} (1969) 669

\bibitem{Coon2} 
M.~Baker, D.~D.~Coon, 
Phys.~Rev. D {\bf 2} (1970) 2349

\bibitem{Diff} 
R.~Schicker, Diffraction 2017,
arXiv:1701.04810

\bibitem{Szanyi} 
L.~Jenkovszky, I.~Szanyi, 
{\it Fine structure of the diffraction cone,} 
arXiv:1701.01269

\bibitem{Barut} 
A.~O.~Barut, D.~E.~Zwanziger, 
Phys.~Rev. {\bf 127} (1962) 974

\bibitem{Statboot} 
S.~Frautchi, 
Phys.~Rev. D {\bf 3} (1971) 2821

\bibitem{StatbootA}
W.~Nahm, 
Nucl.~Phys. B {\bf 45} (1972) 525

\bibitem{Entropy16}
T.~S.~Bir\'o, P.~V\'an, G.~G.~Barnaf\"oldi, K.~\"Urm\"ossy,
Entropy {\bf 16} (2014) 6497.

\bibitem{SQM16}
T.~S.~Bir\'o, G.~G.~Barnaf\"oldi, G.~B\'{\i}r\'o, K.~M.~Shen,
J.~Phys.~Conf.~Ser. {\bf 779} (2017) 012081.

\bibitem{BiroNeda}
T.~S.~Bir\'o, Z.~N\'eda,
Phys.~Rev. E {\bf 95} (2017) 032130


\end{thebibliography}
\end{document}